\def\deg{^{\circ}}
\def\etal{{\it et al.}\thinspace}

\def\eg{{\it e.g.}\thinspace}
\documentclass[useAMS,usenatbib,letterpaper]{mn2e}
\usepackage[totalwidth=480pt, totalheight=680pt]{geometry}
\usepackage{graphicx}

\title[Bistable Profile Illumination in B0919+06 \& B1859+07]{Bistable Profile Illumination in Pulsars B0919+06 \& B1859+07}
\author[Joanna M. Rankin$^{1}$, Cameron Rodriguez$^{1}$ \& Geoffrey Wright$^{2}$ ]
{Joanna M. Rankin\thanks{Joanna.Rankin@uvm.edu}, Cameron Rodriguez and Geoffrey A. E. Wright\\
$^{1}$Physics Department, 405 Cook Physical Science Building, University of Vermont, Burlington, 05405, USA\\
$^{2}$Astronomy Centre, University of Sussex, Falmer, BN1 9QJ, UK : gae@sussex.ac.uk}

\begin{document}

\date{Accepted 2005 month day. Received 2005 month dat; in original form 2005 month day}

\pagerange{\pageref{firstpage}--\pageref{lastpage}} \pubyear{2004}

\maketitle

\label{firstpage}

\begin{abstract}
A new single-pulse behaviour has been identified in two pulsars, B0919+06 
and B1859+07.  Normally both stars emit bright subpulses in a region near 
the trailing edge of their profile.  However, occasionally these stars exhibit 
``events'' wherein the emission longitude gradually decreases, by about 
their profile width, remains in this position for typically several tens of pulses, 
and then gradually returns over a few pulses to the usual longitude.  The 
effect bears some resemblance to a profile ``mode change'', but here the 
effect is gradual and episodic.  When the separate profiles of the normal and ``event'' emission
 are studied, they reveal a broad and complex 
profile structure in each pulsar---but one which can probably be 
understood correctly in terms of the geometry of a single conical beam. Possibly the 
effect entails an extreme example of ``absorption''-induced profile asymmetry, as suspected in other pulsars. Alternatively, shifting sources of illumination within the pulsar's beam may be responsible.  
\end{abstract}


\section{Introduction}
Pulsar emission phenomena have proven to be a rich area of study, 
and these specific effects have often provided needed insights for 
theory building.  The "classical" six such effects, drifting subpulses, 
nulling, profile mode changing, microstructure, polarization modes 
and "absorption" are well known, and several further well defined 
behaviours have been identified in recent years.

We describe below a further such effect which has now been found 
in two poorly studied pulsars in the course of analysing sensitive new 
Arecibo polarimetric pulse-sequence (hereafter PS) observations.  The 
two stars, B0919+06 and B1859+07, both exhibit highly asymmetric 
average profiles with a long leading region of weak emission culminating 
in a bright "component" with a sharp trailing edge.  What attracted our 
attention, however, was that their bright subpulses, usually found in 
the trailing region, occasionally moved progressively over near the 
leading profile edge for several score pulses and then progressively 
back again.  This effect -- which we denote as an "emission shift" -- 
is not a drift, nor is it strictly a profile mode change.  Nor is it an effect 
easily visualized in terms of a rotating subbeam ``carousel'' sytem.  

It is possible that the effect is related to the increasingly fascinating and 
potentially important property of "absorption", or even some more exotic ``Christmas lights'' 
phenomenon whereby different parts of the profile illumine at different times. 
Therefore, we thought it important to present a preliminary study of the phenomenon so that 
other pulsar investigators may become aware of it. \S 2 describes our observations, \S 3 \& 4 examine the character 
of the phenomenon from the perspectives of individual pulses and 
average profile, respectively.  \S 5  \& 7 assess the implied emission 
geometries of the two stars, \S 7 entertains possible explanations and 
\S 8 summarizes the results.  

\begin{table}
  \caption{Available Observations}
  \begin{tabular}{ccccc}
  \hline
    Frequency  &  MJD & Resolution  & pulses & events \\
         (MHz)     &            & ($\deg$)     &               &   \\
  \hline
  B0919+06  &&& \\
  1404 & 44857 &  0.42 & 3582 &      2    \\
  1404 & 44859 &  0.84 & 1886 & none  \\
  1400 & 52854 &  0.43 & 1115 &      2    \\
    327 & 52916 &  0.43 & 4180 &      1    \\
  B1859+07  &&& \\
  1400 & 52739 & 0.64 &  1021 &      6    \\
  1400 & 53372 & 0.64 &  2096 &   8-10 \\
\hline
\end{tabular}
\end{table}

\begin{figure*} 
\includegraphics[width=80mm]{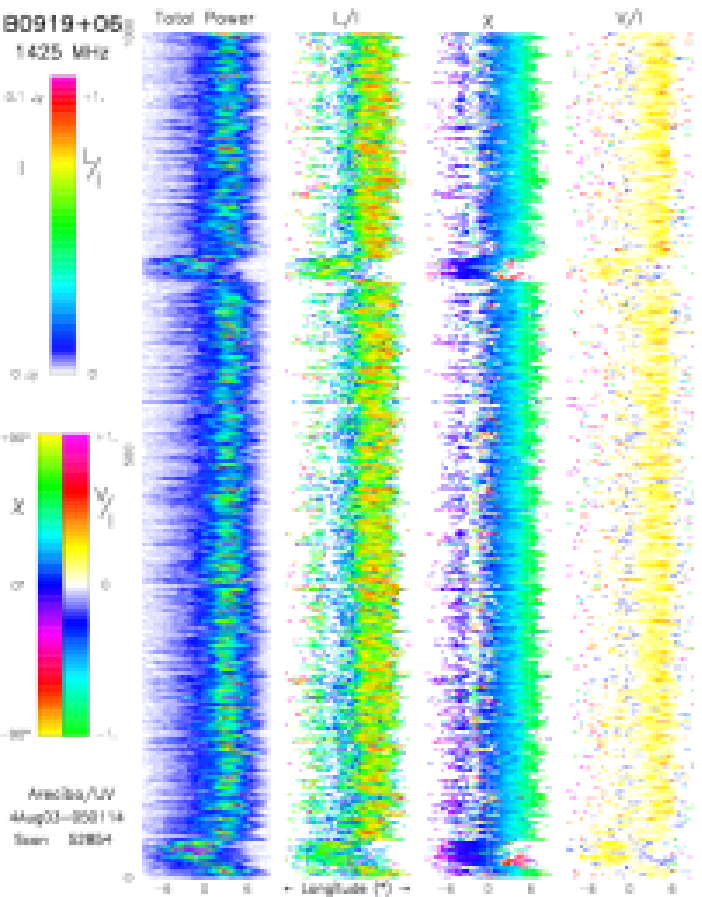}
\includegraphics[width=80mm]{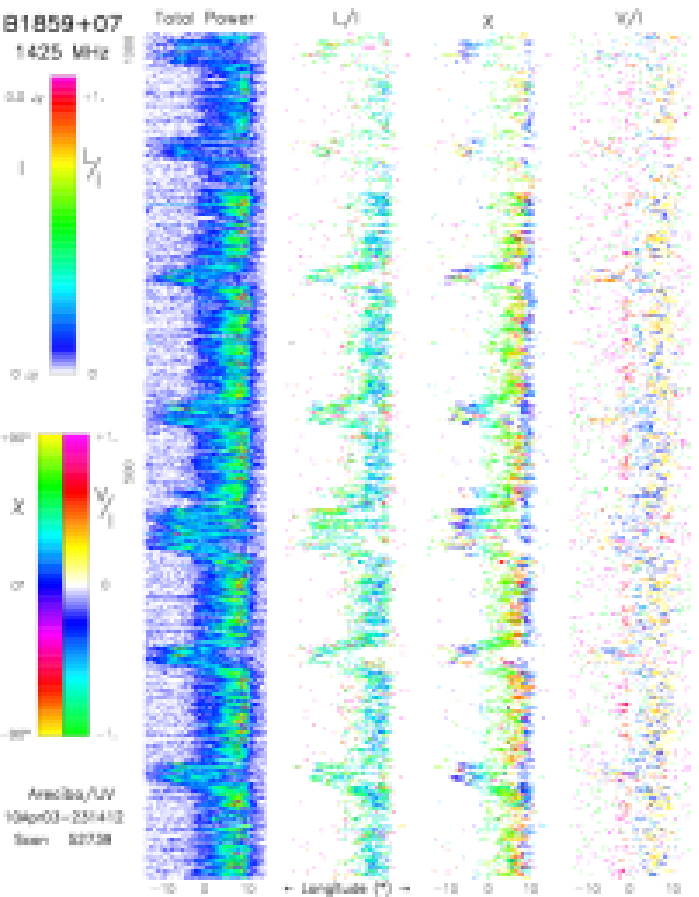}
\caption{Pulse-sequence polarization displays showing ''emission shift'' effect.  
Two such events are seen in the B0919+06 sequence (left) and some 
seven in the B1859+07 (right) display.  In both cases the onset and return 
usually occurs over a duration of a few pulses with bright subpulses appearing 
to shift from the usual emission pattern to an earlier phase.  Both 
panels display a 1000-pulse sequence.  The total power $I$, fractional 
linear $L/I$, PA $\chi$, and fractional circular polarization $V/I$ are 
colour-coded in each of the respective four columns according to their 
respective scales at the left of the diagram.  The latter three columns 
are plotted only when a given sample falls above a 2-sigma noise 
threshold.}
\label{Fig1}
\end{figure*}

\section[]{Observations}
The observations used in our analyses were made using the 305-m 
Arecibo Telescope in Puerto Rico.  The primary L and P-band polarized 
PSs were acquired using the upgraded instrument together with the 
Wideband Arecibo Pulsar Processor (WAPP\footnote{http://www.naic.edu/$\sim$wapp}) 
between 2003 July and early 2005  The ACFs and CCFs of the channel 
voltages produced by receivers connected to orthogonal linearly polarized 
feeds\footnote{A quadrature hybrid was inserted between the feed and  
receivers of the P-band system on 11 October 2004, making it thereafter  
an orthogonal circular system.} were 3-level sampled.  Upon Fourier 
transforming, up to 256 channels were synthesized across a 100- (L) or 
25-MHz (P) bandpass with a sampling time of roughly a milliperiod, 
providing overall resolutions well less than 1$\deg$ longitude.  The 
Stokes parameters have been corrected for dispersion, interstellar 
Faraday rotation, and various instrumental polarization effects.  The 
L-band observations usually recorded four 100-MHz channels centred 
at 1275, 1425, 1525 and 1625 MHz, and the lower three of these were 
added together appropriately for greater sensitivity.  Older Arecibo 
observations at 1400 MHz (Stinebring \etal\ 1984) were used for 
comparison as detailed in Table 1.

\section[]{The Pulse-Sequence Phenomenon}
Figure~\ref{Fig1} displays two 1000-pulse sequences (hereafter PSs), one 
of B0919+06 (left) and the other of B1859+07 (right).  Both exhibit episodes 
in which the emission moves sharply earlier leaving the usual region of 
bright subpulses almost empty.  Two such ``events'' are seen in B0919+06 
and some 7 in B1859+07.  In both stars the emission moves earlier by an 
amount which is of the scale of the profile width.  Each event a) commences 
and exhibits an orderly shift, over a few pulses, of bright emission toward 
earlier longitude, b) remains in this state for up to some 20 or so pulses, 
and c) then gradually returns over a few pulses to the usual configuration 
and phase of emission.  In B0919+06, theses events are rather rare---one 
typically in several thousand pulses---which may account for why they have 
not been reported earlier.  In B1859+07, however, they typically occur every 
1--2 hundred pulses, and in Fig.~\ref{Fig1} they even appear quasi-periodic 
though much less so in the MJD 53372 observations (not shown).  

In total power, the events almost resemble an effect which could be produced 
by an instrumental timing fault, and perhaps this is another reason that they 
have not been noted earlier.  In polarization they show up more clearly, in 
part because the linear and circular is plotted only when it exceeds a noise 
threshold.  This 2-sigma threshold is shown as a narrow white bar on the $I$-$L/I$
color bar, but it can only be seen on the B1859+07 bar at about the 5\% level 
because in B0919+06 the signal-to-noise ratio (hereafter S/N) is so high that 
it disappears into the white region just above zero intensity.  

\begin{figure} 
\includegraphics[width=80mm]{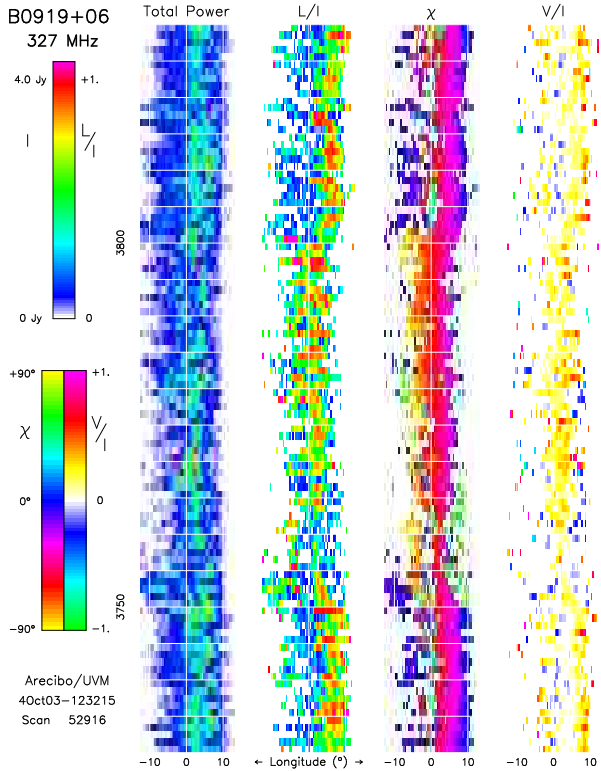}
\caption{Pulse-sequence polarization display showing a ''shift'' effect  
in B0919+06 at 327 MHz as in Fig.~\ref{Fig1}.  Here only a 200-pulse 
sequence is depicted.  The event begins about pulse 3750 and extends 
to pulse 3710.}
\label{Fig2}
\end{figure}

Close inspection shows that the events have different and characteristic linear 
polarization angles (hereafter PAs):  In B0919+06 the PA rotates smoothly 
from green ($+60\deg$) on the trailing edge through cyan ($+30\deg$) to 
almost blue ($+15\deg$) on the leading edge of the usual profile, but during an 
event the PA reaches full purple ($-15\deg$).  Similarly, in B1859+07 note that 
there is an region of smooth PA rotation that begins with near orange ($-75\deg$) 
and rotates through green ($+60\deg$) to cyan ($+30\deg$) and even blue 
($0\deg$) during an event.  Also, there is a PA jump on the trailing edge to 
blue ($0\deg$).  

In B0919+06's normal emission pattern note that there is a region of enhanced 
linear polarization between about --3 and --5\degr longitude and further that 
during the two events bright subpulses extend well in advance of --6\degr.  It 
is also worth noticing that the region of reduced linear polarization on the 
leading edge of the normal pattern---which we have taken as the longitude 
origin---also tends to occur on the trailing side of the event pattern.  And both 
the trailing part of the normal pattern and leading side of the events show 
some weak negative circular polarization.  B1859+07's circular polarization 
is more interesting: we see consistent weak positive at about --2\degr and a 
suggestion of weak negative circular around +2\degr longitude.  

Figure~\ref{Fig2} shows an expanded view of the one event encountered in 
our only 327-MHz observation of B0919+06 on MJD 52916.  At this lower 
frequency the overall profile width is larger so that the event is less dramatic.
Interestingly, it is hardly discernible in in total power, but in the linear and even 
circular polarization we see that the band of highly linearly polarized subpulses 
moves sharply to earlier longitudes following pulse number 3750 and remains 
in this configuration until or just after pulse 3800.  The shift is somewhat more 
apparent in the three polarization columns because the noise threshold defines 
the leading and trailing edges of the plotted area.  Note, though, that the PA 
values during the event represent a smooth rotation toward the yellow-orange 
angle ($-80\deg$) from the blue-magenta-red (0 to $-60\deg$) PA range that 
is characteristic of the usual emission pattern.

\begin{figure*}
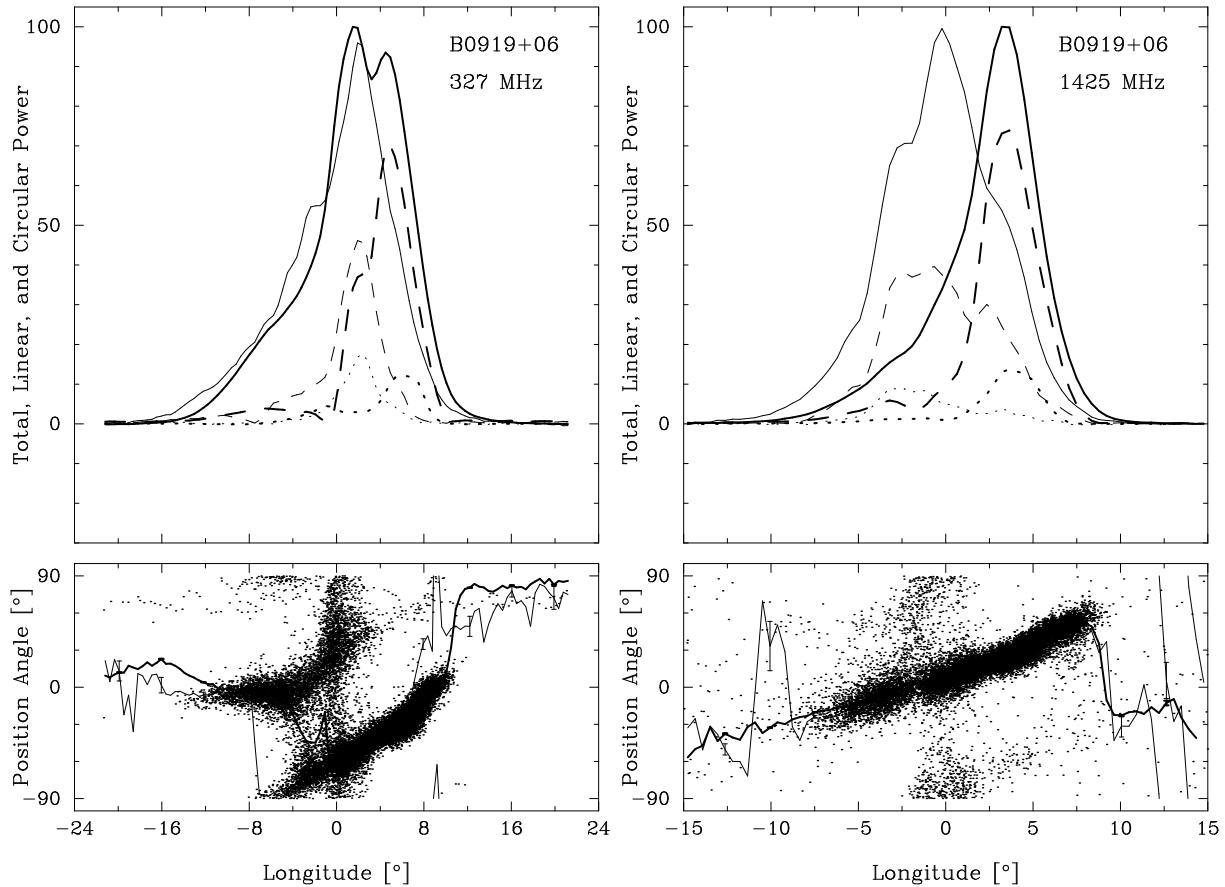
 
\includegraphics[width=80mm]{fig3a.ps}
\includegraphics[width=80mm]{fig3b.ps}
\caption{Total (bold) and partial average (lighter curves) profiles of B0919+06 
at 327 MHz on MJD 52916 (left) and 1400 MHz on MJD 52854 (right) in the 
upper panels.  Solid curves are the total intensity (Stokes $I$), dashed the 
total linear (Stokes $L$), and dotted the circular polarization (Stokes $V$).  
The lower panels give PA histograms corresponding to the total profile of 
those samples having PA errors smaller than 3\degr as well as the average 
PA curves for both the total and partial profiles.  The total profiles are very 
asymmetric with gradual leading and sharp trailing edges.  The partial averages 
include only those pulses associated with the ``events'' in the respective PSs.  
Note the markedly different profile forms but highly similar PA behavior.}
\label{Fig3}
\end{figure*}

\begin{figure*}
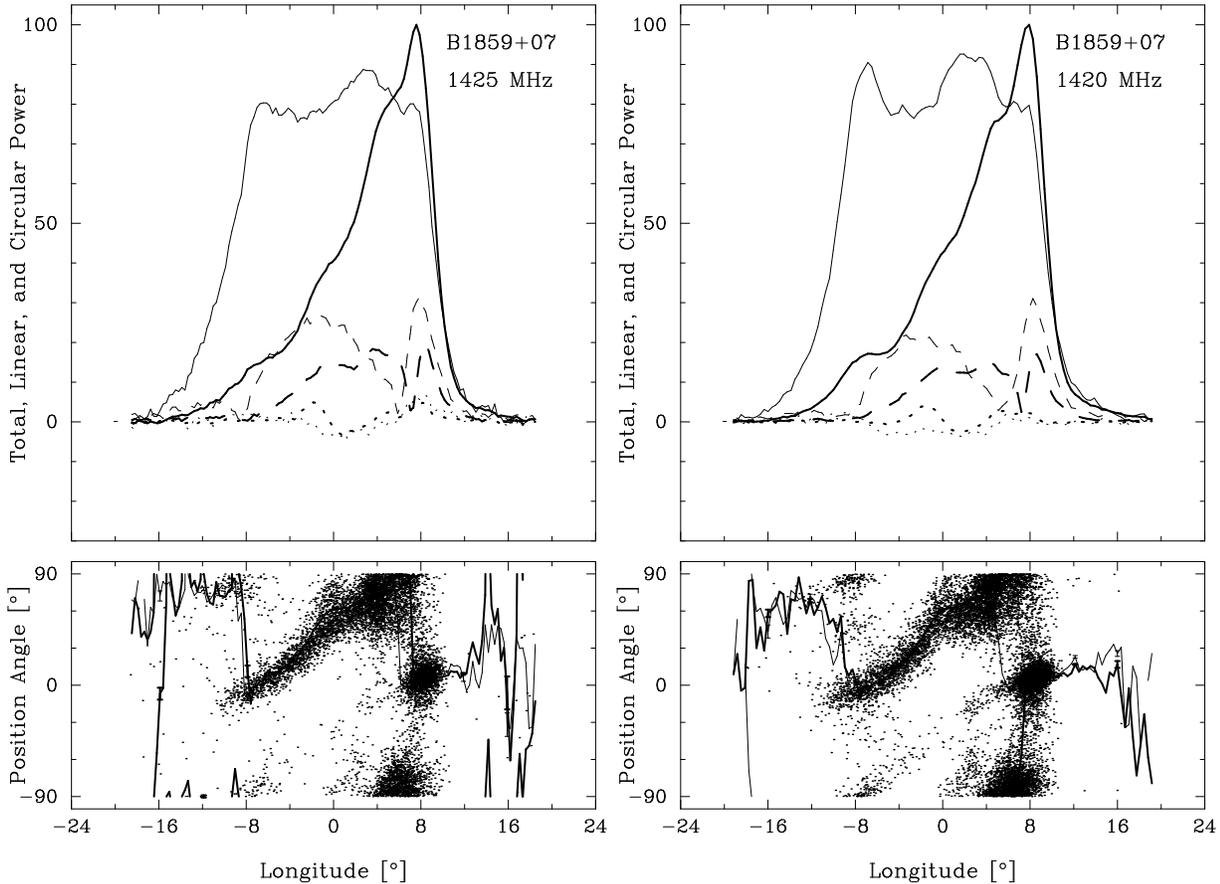
 
\includegraphics[width=80mm]{fig4a.ps}
\includegraphics[width=80mm]{fig4b.ps}
\caption{Total (bold) and partial average (lighter curves) profiles of B1859+07 
at 1400 MHz on MJD 52739 (left) and 53372 (right) as in Fig.~\ref{Fig3}.  Here 
the partial averages are scaled up by a factor of 1.5 for easier comparison 
with the total profiles.  The lower panels give PA histograms of those samples 
having PA errors smaller than 3\degr as well as the average PA curves for 
both the total and partial profiles.  The total profiles are again asymmetric 
with gradual leading and sharp trailing edges, whereas the partial averages 
including only the event pulses are nearly symmetric.  Again note the markedly 
different profile forms but highly similar PA behavior.}
\label{Fig4}
\end{figure*}

\section[]{Average Profile \& Modulation Properties}
Let us now try to understand the ``shift'' events further by constructing sets 
of total (bold) and partial average (lighter curves) profiles.  These are shown in 
Figures~\ref{Fig3} for B0919+06 and \ref{Fig4} for B1859+07.  The top panels 
give the total intensity (solid curves), total linear polarization (dashed lines) and 
circular polarization (dotted lines).  The lower panels show the total PA distribution 
of samples having angle errors less than 3\degr as well as average PA curves 
of the total (bold) and partial (lighter) averages.  

The asymmetric total profiles of B0919+06 are well known from Stinebring \etal\ 
(1984) and Blaskiewicz \etal\ (1991), and even in the latter it was understood 
that the single bright peak at  21 cms aligns with the trailing component at 430 
MHz, probably owing to the work of Phillips \& Wolszczan (1991).  Fig.~\ref{Fig3} 
shows us that the star's profiles (bold curves) have three main components at 
both frequencies.  At 327 MHz (left diagram) the center and trailing components 
are almost equally bright with the leading one much less so.  During an event, 
however (lighter curves), the leading component brightens up, the central one 
remains almost unchanged, and the trailing feature's intensity drops to half or 
less.  Note that the strong linear polarization under the trailing feature simply 
disappears during the event.  Then we see a polarized feature associated with 
the central component, a leading-edge extension of the region of smooth PA 
rotation, and indeed a poorly defined linear PA in the region under the trailing 
component. 

A similar, but more dramatic behavior is seen at 1400 MHz in the right-hand 
display (Note the finer longitude scale).  Here the total profile asymmetry is so 
extreme that we see little apart from the trailing component.  The central feature 
is visible in neither the total power nor the linear polarization, and the leading 
component shows only as a small bump in both.  During the events, however, 
(lighter curves) the profile position shifts markedly earlier, and the three main 
components are in clear evidence.  The leading component is again much 
brighter, the central feature now as bright as the usual trailing feature, and the 
latter's intensity now again reduced to half or so.  We also see that the PA 
traverses of the total and event profiles track each other closely.  Finally, recall 
that the events in this pulsar are rather rare as shown in Table 1, so that the 
partial averages at 327 and 1400 MHz are comprised of only some 36 and 
77 pulses, respectively.

Figure~\ref{Fig4} gives a further set of total and ``event'' partial average profiles 
for the two observations of B1859+07 at 1400 MHz on MJDs 52739 and 53372.  
This star's two total average profiles (bold curves) are also very asymmetric with 
long gradual leading regions and sharp trailing edges.  Both observations also 
show several weak features which may be evidence of underlying component 
structure.  Then, during the events---which are considerably more frequent in 
this pulsar---the profiles (lighter curves) become more nearly symmetric.  

While the total profiles do not exhibit any recognizable profile form, the partial 
averages are much more interesting.  Notice first the tripartite form of the total linear 
polarization $L$: we see a trailing feature aligned with the usual bright narrow 
total-power trailing component at about +9\degr longitude, then a broad central 
``hump'' between about $-8$ and +7\degr, and finally a leading feature around 
$-10\deg$.  The latter's linear power is small, but note in the lower panel that 
two different populations of bright polarized subpulse samples contribute to the 
power in this leading-edge region---one at a PA of roughly $-10\deg$ and the 
other at some +80\degr---contributing to the nearly complete depolarization on 
the leading edge of the partial profile.  We can also see that this polarization-modal 
behavior on the leading edge mirrors that on the trailing edge, which there has 
primary polarization-mode power at about +90\degr immediately followed by 
secondary polarization-mode power at about 0\degr.  Such adjacent configurations 
of modal power---producing 90\degr PA ``flips'' and linear polarization ``nulls''---are 
most often seen on the outside edges of conal beams (Rankin \& Ramachandran 
2003).  Thus, they seem to mark the sightline crossings of the outer conal beam 
edge and indicate that even this extremely asymmetric profile is produced in 
substantial part by a full traverse through a conal emission beam.  Again, it is 
worth noting that the PA distribution reflects the properties of the total profile, but 
of course emission at longitudes earlier that about $-5\deg$ only occurs during 
the ``events'' as can readily be seen in the right-hand panel of Fig.~\ref{Fig1}.  

Moreover, we see weak antisymmetric circular polarization in the total profiles 
about the longitude of the origin, and while it is only a few percent positive (near 
$-2\deg$) and negative (near +2\degr) it is produced by a population of highly 
circularly polarized subpulses as can again be seen in the righthand panel of 
Fig.~\ref{Fig1}.  This in turn suggests some core activity in the center part of the 
profile and argues that overall we should regard the star's profile as reflecting 
emission from both a core and a cone.   

Finally, Weltevrede \etal\ (2005) include both stars in their survey of pulse 
modulation characteristics.  In the case of B0919+06, not surprizingly they find 
evidence of a weak modulation with a $P_3$ of 50 $P_1$ or more.  In the case 
of B1859+07, their fluctuation spectra are almost ``white'', perhaps because the 
active leading-edge region was too weak to be detected.

\section[]{B0919+06 Profile Geometry}
\begin{figure} 
\includegraphics[width=80mm]{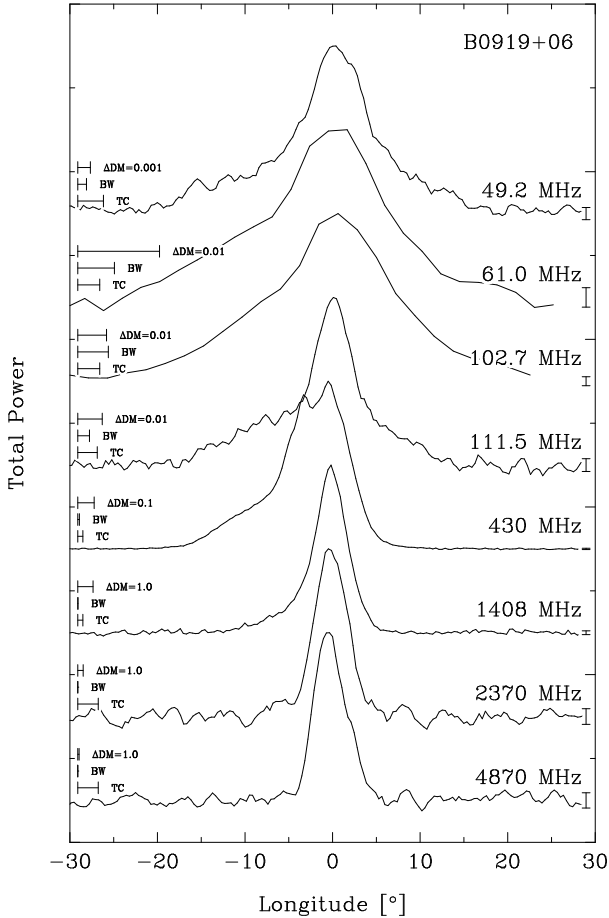}
\caption{Time-aligned average profiles of B0919+06 between 49 and 4870 MHz. 
All were recorded using the Arecibo Observatory but those at 61 and 103 MHz 
which acquired at the Pushchino Radio Astronomy Observatory.  Bars indicating 
the instrumental resolutions are given at the left of the diagram.  From Hankins 
\& Rankin (2006).}
\label{Fig5}
\end{figure}

Pulsar B0919+06 is bright over a very large spectral range and has thus been 
well observed both at very high and very low frequencies.  At 21 cms and above, 
its profile exhibits only a single component---the one we have identified as its 
trailing component---as we have seen in Fig.~\ref{Fig3} (right).  Then, in a range 
including 300--500 MHz, it shows two components (\eg, Fig.3, left), spaced by 
about 3\degr.  At all lower frequencies the resolution has been such that these 
two features apparently merge into one broad feature.  The entire sweep of this 
star's profile evolution can be seen in Figure~\ref{Fig5}, where a set of high 
quality total power profiles have been time-aligned over a frequency range from 
50 to 5000 MHz (from Hankins \& Rankin 2006).  

Of course, these profiles should not have been aligned as they are:  The star's 
profiles are basically three-componented, so the middle (probably core) feature 
would provide a better alignment point.  This would shift the longitude origin 
some 4\degr earlier---near the leading peak of the 430-MHz profile---so that the 
second peak as well as the higher frequency profile peaks align as a trailing 
conal component.  Then, the time-alignment of the 100-MHz and lower profiles 
seems to indicate that they too are dominated by this trailing component, and 
in no other profile do we clearly see the central component.  

Quality polarimetry is available for B0919+06 at many frequencies from a 
number of different instruments.  In addition to Blaskiewicz \etal\ and Hankins 
\& Rankin above, Gould \& Lyne (1998), von Hoensbroech \& Xilouris (1997), 
Rankin \etal\ (1989), Stinebring \etal\ (1984), Suleymanova \etal\ (1988) and 
Weisberg \etal\ (1999) have published profiles between 60 MHz and 5 GHz.  
Unfortunately, most of this work gives little insight into the leading ``ramp'' 
region of most interest to us here.

Another hint may come from the size of the triangular ``ramp'' on the profile 
leading edges.  We first see evidence of this in the asymmetry of the 1408-MHz 
profile, which we know from the above discussion is associated with the ``events'' 
and therefore with activity in the earlier components.  Here, its overall scale 
(see also Fig.~\ref{Fig3}) is hardly 10\degr, but from 430 MHz down it appears 
to have a nearly constant extent of about twice this.

The nearly symmetrical triangular ``ramp'' on this star's trailing edge at low 
frequencies is no less strange.  We see no hint of it in our 327-MHz profile, 
but at 111 MHz, it is a significant feature.  Nor can this be attributed to scattering; 
Phillips \& Wolszczan (1992) have shown that scattering dominates the star's 
profile at 25 MHz, but appears to distort it little at 50 MHz.  Moreover, the 
available polarimetry at low frequencies (Suleymanova \etal\ 1988; Hankins 
\& Rankin 2006) suggest that the triangular ``ramps'' are highly linearly polarized 
at an angle rotating about +9\degr/\degr over the entire profile.  These triangular 
low frequency features---not really components---are quite unusual and deserve 
further study.

Finally, returning to Fig.~\ref{Fig3} the partial profiles provide further indications 
of the profile's structure, but not enough to significantly improve the geometrical 
models in Paper VIb.  Slightly revising the values there (see Table 5), a putative 
core width of some 4.7\degr,  1-GHz profile width of about 10\degr, and a PA rate 
of +9\degr/\degr indicates an inner cone geometry with magnetic latitude and 
sightline impact angles of 53 and 5.1\degr, respectively, and a sightline cut some 
25\% inward of the outside conal 3-db edge.  This may be compatible with the 
apparent lack of conal spreading at very low frequency.  

\section[]{B1859+07 Profile Geometry}
A more limited record of published observations is available for B1859+07, in 
part because its relatively short period (0.644 s) combined with a large dispersion measure (253 
pc/cm$^3$) make it difficult to observe at lower frequencies.  Polarimetric 
observations have been published only by Gould \& Lyne (1998) and Weisberg 
\etal\ (2004).\footnote{Both observations show anomalies at lower frequencies: 
The former's 300-MHz profile is poorly resolved and defined polarimetrically; 
whereas the latter authors were unable to detect significant linear polarization.}  
Both show single profiles with a slow rise and steep trailing edge.  The average 
profile width increases with decreasing frequency; its half-power dimension is 
hardly 10\degr\ around 21 cm and may be as much as 20\degr\ in the 400-MHz 
region.  

At no frequency does the star exhibit much fractional linear linear polarization, 20\% 
at most.  Gould \& Lyne's 1408-MHz profile resembles those in Fig.~\ref{Fig4} in 
both form and fractional linear; even the PA seems to show both the leading ramp 
and abrupt $-90\deg$ ``jump''.   The PA rate is about +6\degr/\degr, and as we had 
observed above, the clear ``patches'' of modal polarization on the edges of the 
event partial profile, together with the evidence for low frequency spreading, have 
the pattern of an outer cone.  Moreover, the emission near the center of the profile 
suggests core activity---and altogether argues that the underlying emission pattern 
is that of a triple ({\bf T}) or possibly a five-component ({\bf M}) geometry.  

We can then compute the basic emission geometry as above using the procedures 
outlined in Rankin (1993b) for Table 5.  Estimating the outside (conal) half-power 
1-GHz width of the profile as 20\degr\ from the ``event'' profiles in Fig.~\ref{Fig4}, 
we find that the star has an outer cone geometry with $\alpha$ and $\beta$ some 
31\degr\ and 4.8\degr, respectively.  Interestingly, this geometry also implies a 
core width of 6.0\degr, and while we are not able to measure this width accurately 
from the profiles, we can get some indication of its value from the circularly polarized 
power in individual pulses referred to above.  If we measure the typical separation 
between the respective leading positively and weaker trailing negatively circularly 
polarized subpulses associated with the core emission in Fig.~\ref{Fig1} (\eg, near 
pulse 150 or 650), the interval is clearly a few degrees.  In total power the central 
feature in the ``event'' partial profiles of Fig.~\ref{Fig4} appears quite broad, but 
could also be comprised of several merged components (if intrinsically a double 
cone {\bf M} profile).

\section[]{How are ``Events'' to be Understood?}
Since first noting the ``emission shift'' phenomenon a few months ago, we have tried 
to understand its character and causes.  We considered, for instance, whether 
the ``events'' might be produced by a displacement of the emitting region to 
a higher altitude within the polar flux tube.  This idea at first seemed appealing 
given the gradual nature of the ``event'' onsets and returns. However,  
on exploring the idea quantitatively, the displacements seemed too large in terms of 
vertical height in the magnetosphere. In both pulsars the magnitude of the swing is 
about 15 degrees, and interpreted in terms of light-travel-time would correspond to a 
vertical displacement about one quarter of the light-cylinder radius (5,500km in the case of B0919+06
and 8,000km for B1859+07). This figure greatly exceeds the height above the 
neutron star's surface within which the radio emission is widely believed to originate, 
and must be considered implausible.
Furthermore, we have demonstrated in both pulsars a consistency with the picture of 
a single emission cone for both normal and shifted emission, based on the continuous PA
change across the profile. A sudden increase in emission altitude would surely lead to an abrupt change
to some new PA pattern and a dramatic widening of the total emission cone, 
and neither of these features are observed.  

Rather, on closer inspection, the usual profile seems to reflect an incomplete 
illumination of the full geometric traverse of the sightline through the emission 
cone.  During the ``events'', the situation is very different:  here the partial profiles 
have recognizable forms, dimensions, component structures and even sensible 
classifications (\eg, Rankin 1983a).  In both cases, the usual total profile is 
unusually asymmetric, but the partial profiles of the ``events'' have the usual 
and familar level of overall conal symmetry.  Similarly with the PA behaviour, 
especially of B1859+07:  the PA histograms suggest a complete profile with 
modal power on both edges, but this power is only observed on the leading 
edge during ``events''. 

Given these circumstances, we began to ask whether the ``events'' could be 
constructively viewed as a kind of profile ``mode change''.  Surely we do see 
a good deal of evidence for more or less asymmetric profile illumination in the 
different modes of a given star.  What makes the ``emission shifts'' stand out, though, 
is their gradual character:  usually we see that the stars take a few single pulses 
to go from the ``normal'' emission pattern to that of an ``event''---and in general 
mode changes seem to occur on a time scale of a single pulse.  We can think 
of no good case of a gradual mode-change onset.\footnote{We do have some 
evidence for a gradual as well as sharp character to mode changes in pulsar 
B0943+10 (Suleymanova \etal\ 1998; see also Rankin \& Suleymanova 2005)} 
A recognizable reconfiguration of the emission pattern within about about a 
single stellar rotation is thus usually implied in a ``mode change''.  

If the ``events'' are not a form of ``mode change'' (or even if they are), we have 
seen above that in both pulsars they entail a change in the emission pattern, 
from one which is concentrated ``late'' on the trailing edge of the profile---that 
is well past the probable longitude of the magnetic axis---to one which strongly 
favors a region earlier than the zero longitude.  Such a configuration and effect 
is very suggestive of the problematic ``absorption'' phenomenon first identified 
in pulsar B0809+74 (Bartel \etal, Bartel 1981), but now identified in a number 
of situations (\eg, Rankin 1983b).  Indeed, in the current context ``absorption''
provides an interesting model:  if one looks carefully again at Fig.~\ref{Fig1}, it 
{\em does} appear that the leading part of the emission pattern is usually 
obscured, but during an ``event'' it is rather the trailing part which is obscured.  

A good deal of evidence in other pulsar contexts does indeed suggest that a 
partial or complete blockage of the radiation from either the leading of trailing 
side of the magnetic axis longitude does occur.  The most interesting recent evidence about 
the phenomenon comes from studies of the drifting-subpulse patterns of pulsars 
B0943+10 (Deshpande \& Rankin 1999, 2001; Rankin \etal\ 2003; Rankin \& 
Suleymanova 2005) and B0809+74 (Rankin \etal\ 2005a,b,c), where the 
rotating-subbeam ``carousel'' is usually visible only at longitudes later and earlier 
than the central longitude, respectively.  Fascinatingly, however, this is not always 
the case: in B0809+74 the ``absorption'' characteristics are frequency dependent 
and possibly are in some cases specific to one polarization mode.  In B0943+10 
recent studies show that the ``absorption'' properties can change very slowly; the 
profile changes entailed in transitions from the star's Q to B emission modes take 
of order an hour to completely change from favouring emission prior to the central 
longitude to nearly precluding it.

However, here we have deliberately and cautiously maintained inverted commas around the word ``absorption'', 
since the presence of an absorbing agent between ourselves and the source may not be the only
interpretation for the emission shifts in B0919+06 and B1850+07. We might equally 
suppose that different regions of the open emission cone defined by the profile structure illumine
at different times, in the manner of ``Christmas lights'', so that the emission flips 
from one side of the magnetic axis to the other. However, both explanations demand 
dynamic, irregular changes in the pulsars' magnetospheres--whether in the absorbing or emitting material--
 for which no model has yet appeared, and there would seem to be no observational 
 way to distinguish between the two hypotheses.

An interesting point is that both B0919+06 and B1859+07 usually emit late and the ``events''
shift the emission earlier.  Other prominent instances of ``absorption'' show no 
such asymmetry: as noted above B0943+10's effect usually curtails emission 
after the central longitude, whereas just the opposite is true for B0809+74.  Thus, 
we cannot be sure whether there might be other pulsars which could usually emit 
early and then throw their emission later during ``events''.  Indeed, an in some ways 
very similar effect is seen in B2034+19, but here the changes are quick, bounded 
by nulls and are easily classed under the ``mode-change'' rubric (Redman 2006).  It 
is also possible that in some stars the ``events'' occur much more frequently so that 
they are seen merely as chaotic subpulse modulation;  possibly pulsar B1604--00 
provides such an instance (\eg, Rankin 1988).

Taking the pulsars B0919+06, B1859+07 and B1604-00 as the most convincing exemplars of the 
emission shift phenomenon, we attempted to find some common features among their basic
physical parameters. Certainly, their spin-down ages give no clue: they range from the 
``middle-aged'' B0919+06 at 0.5Myr to B1604-00 at 20Myr. Much the same is true of their
inferred surface magnetic fields and their corresponding light-cylinder values. The only possible hint is that
their rotation periods are fairly close to one another (0.43s, 0.64s and 0.43s respectively) and somewhat shorter 
than the average for the general population of ``slow'' pulsars. Physically, this may suggest that
the observed $\it {gradual}$ emission shift requires an intermediate-size light-cylinder for its operation.    

\section{Summary}
 We identify what seems to be a new aspect of pulsar behavior wherein the 
 emission usually comes from a trailing region of the profile, but then occasionally 
 shifts earlier over a few pulses to illuminate the leading part of the profile for some 
 20--50 pulses, and then shifts back again over a few pulses.  Both B0919+06 and 
 B1859+07 exhibit the effect, the former perhaps once in 2000 pulses and the 
 latter about 10 times more often.
 
 Both stars exhibit asymmetric single profiles which defy ready classification, 
 though in the case of B0919+06 enough structure has been discerned to 
 suggest that it has three main components (Rankin 1993b).  However, when 
 partial profiles are constructed only of the pulses framing the ``events'', more 
 recognizable profile forms are found---and in both cases we have been able 
 to identify some of the underlying component structure and work out their 
 probable basic emission geometry.  It is interesting that B0919+06 is among 
 the stars thought by Lyne \& Manchester (1988) to represent a class of ``partial 
 cones''---and indeed we certainly find that this classification is appropriate given 
 that only during its ``events'' is the leading part of its cone illuminated.  It will be 
 interesting to see if other pulsars with asymmetric single profiles thus classified 
 by Lyne \& Manchester also exhibit similar ``events''---and whether other stars 
 exhibiting the effect will be found to throw their emission later rather than earlier.  
 
 Our primary interest in these stars and their behaviour, however, lies in the 
 possibility that the ``events'' represent an example of the 
 profile ``absorption'' effect.  In both pulsars it is possible to interpret
 the emission shift as if emission in the leading 
 part of the profile is normally obscured and then, for brief intervals, the 
 obscuration shifts to the trailing part of the star's profiles. ``Absorption'' 
 appears to exhibit this sort of asymmetry about the longitude of the magnetic 
 axis in other pulsars, and it also exhibits a gradual onset in some other 
 instances. However, we may see the effect in terms of shifting illumination within the 
 defined emission cone, so that the emitting zone appears to shift laterally as different 
 regions switch on and off. Clearly, we do not yet understand physically what is the origin of such 
 observed phenomena, but the starting point must be define its observational 
 characteristics.

\section*{Acknowledgments}
We thank Avinash Deshpande for help with the observations.  Portions of this 
work were carried out with support from US National Science Foundation Grant 
AST 99-87654.  Arecibo Observatory is operated by Cornell University under 
contract to the NSF.  This work made use of the NASA ADS system. GAEW thanks 
the University of Sussex for a Visiting Fellowship.

\label{lastpage}

\end{document}